\documentclass{iaus}
\usepackage{graphicx}
\usepackage{lipsum}
\usepackage{natbib}

\title[Surface evolution in stable magnetic fields] 
{Surface evolution in stable magnetic fields: the case of the fully convective dwarf V374\,Peg}

\author[K. Vida, K. Ol\'ah, Zs. K\H{o}v\'ari]   
{K. Vida$^1$, K. Ol\'ah$^1$, Zs. K\H{o}v\'ari$^1$}

\affiliation{$^1$Konkoly Observatory of the Hungarian Academy of Sciences \\
H-1121 Budapest, Konkoly Thege Mikl\'os str. 15-17. 
\\email: {\tt vidakris@konkoly.hu}}

\pubyear{2010}
\volume{273}  
\pagerange{119--126}
\jname{Physics of Sun and star spots}
\editors{Debi Prasad Choudhary\& Klaus G. Strassmeier, eds.}
\begin{document}

\maketitle

\begin{abstract}
We present $BV(RI)_C$ photometric measurements of the dM4-type V374\,Peg covering $\sim430$ days. The star has a mass of $\sim0.28M_\mathrm{Sun}$, so it is supposed to be fully convective. Previous observations detected almost-rigid-body rotation and stable, axisymmetric poloidal magnetic field. 
Our photometric data agree well with this picture, one persistent active nest is found on the stellar surface. Nevertheless, the surface is not static: night-to-night variations and frequent flaring are observed. The flares seem to be concentrated on the brighter part of the surface. The short-time changes of the light curve could indicate emerging flux ropes in the same region, resembling to the active nests on the Sun. 
We have observed flaring and quiet states of V374\,Peg changing on monthly timescale.
\end{abstract}

\firstsection 
\section{Introduction}
V374 Peg is an M4 dwarf \citep{specclass} whose X-ray emission has been detected by the ROSAT satellite \citep{ROSAT}. \cite{ibvs} and \cite{2001AstL.27.29B} presented photometry in $R$ and $UBVRI$ filters, both paper report intense and frequent flares on the star. 
Recently, \cite{NEON} published spectroscopic observations of V374 Peg (along with some of the photometry presented in this paper), which showed intensive emission in the H$\alpha$ line.

A reason why this object is especially interesting is the mass of the star: it is $0.28M_\mathrm{Sun}$ \citep{mass}, which places it just below the theoretical limit of full convection ($0.35M_\mathrm{Sun}$ according to \citealt{convlimit}).
Above this limit stars have a radiative core with a convective envelope and are supposed to sustain a solar-type $\alpha\Omega$ dynamo \citep{parker,babcock,leighton}. Below $0.35M_\mathrm{Sun}$ stars are fully convective, but the origin of their magnetic fields is unclear in details. \cite{dynamo1} and \cite{dynamo2} showed, that these kind of stars can produce large-scale, non-axisymmetric fields using $\alpha^2$-dynamo, if they rotate rigidly. On the other hand, they can have axisymmetric, poloidal fields, given that they have strong differential rotation \citep{dynamo3}. 
As spectropolarimetric observations of \cite{donati} and \cite{morin} have showed V374 Peg has a very weak differential rotation, but at the same time the star has stable, axisymmetric, poloidal magnetic field, which doesn't fit to the current theoretical models.

In this paper we study the behaviour of the photosphere of V374 Peg on a yearly timescale.
\section{Observations}
\begin{figure}[t]
 \centering
 \includegraphics[width=1.0\textwidth]{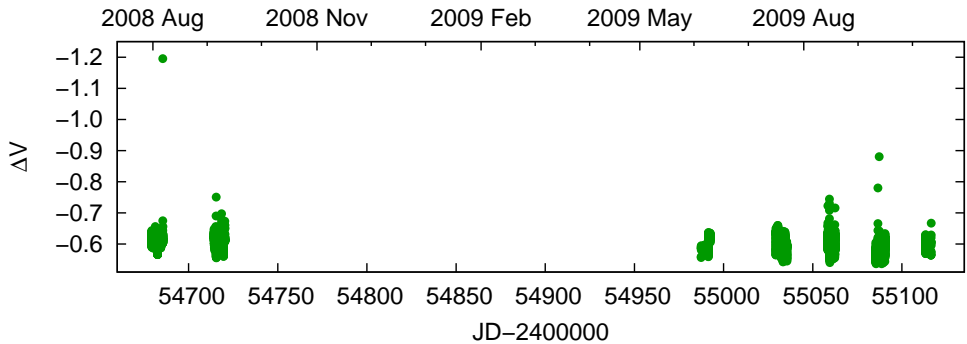}
 \caption{$V$ light curve of V374 Peg. The deviating points are flares.}
 \label{fig:all_v}
\end{figure}

We present photometric data obtained by the 1m RCC telescope of the Konkoly Observatory at Piszk\'estet\"o Mountain Station. The telescope is equipped with a $1300\times1300$ Princeton Instruments CCD. The $BV(RI)_C$ light curves have been obtained between 2008 July 31 and 2009 October 12 in seven observing runs. The observing runs are separated by about 3 weeks. 
Data reduction and aperture photometry was done using standard 
IRAF\footnote{IRAF is distributed by the National Optical Astronomy Observatory, which is operated by the Asso-
ciation of Universities for Research in Astronomy, Inc., under cooperative agreement with the National
Science Foundation.
}
 routines. The resulting light curve in $V$ passband is plotted in Fig. \ref{fig:all_v}.
For the phased light curves we used the following ephemeris:
$${\rm HJD}=2453601.786130+0.445679\times E,$$
where the period was determined for the whole photometric observation using the SLLK method described by \cite{sllk}. This algorithm phases the datasets with different periods and chooses the period giving the 'smoothest' light curve as the correct one, giving a more reliable result than Fourier-analysis for quasi-periodic, non-sinusoidal light curves.

\section{Analysis of the data}

\begin{figure}[t]
 \centering
 \includegraphics[height=9.55cm]{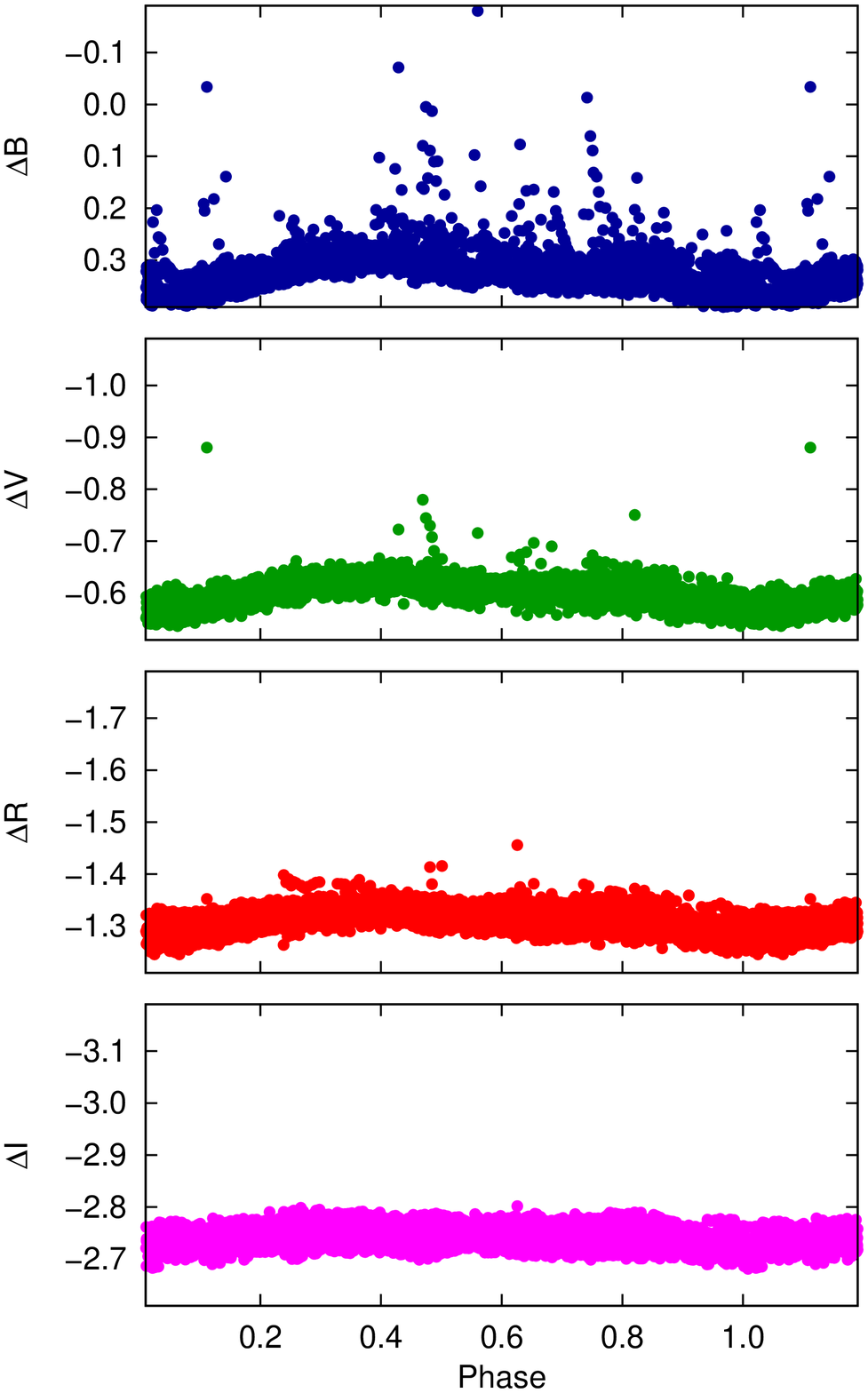}
 \includegraphics[height=9.55cm]{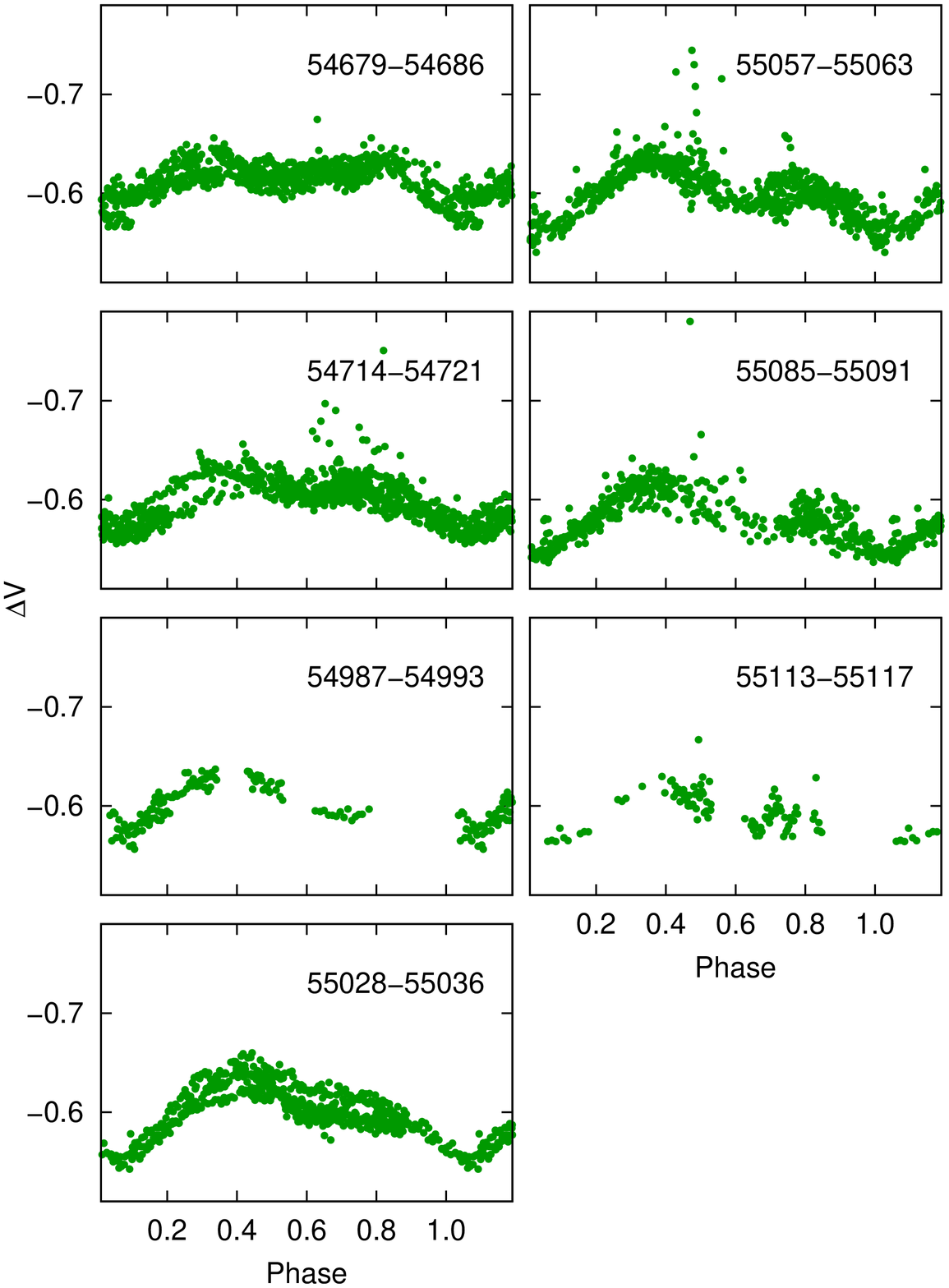}
 \caption{\textit{Left:} Phased $BV(RI)_C$ light curves of V374 Peg showing all the observations. Frequent and intensive flares can be seen in $B$, $V$, and $R_C$ bands. The flares tend to be concentrated around phase 0.5. \textit{Right:} Phased $V$ light curves from different observing runs. The star shows active (flaring) and less active phases changing on monthly timescale.}
 \label{fig:phased}
\end{figure}

\begin{figure}[t]
 \centering

 \includegraphics[width=0.72\textwidth]{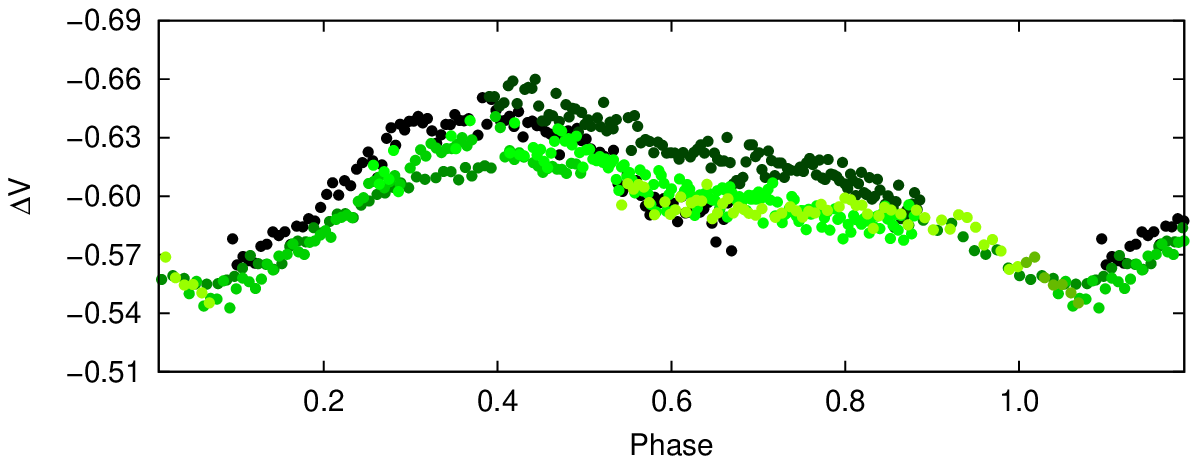}
\includegraphics[width=0.27\textwidth]{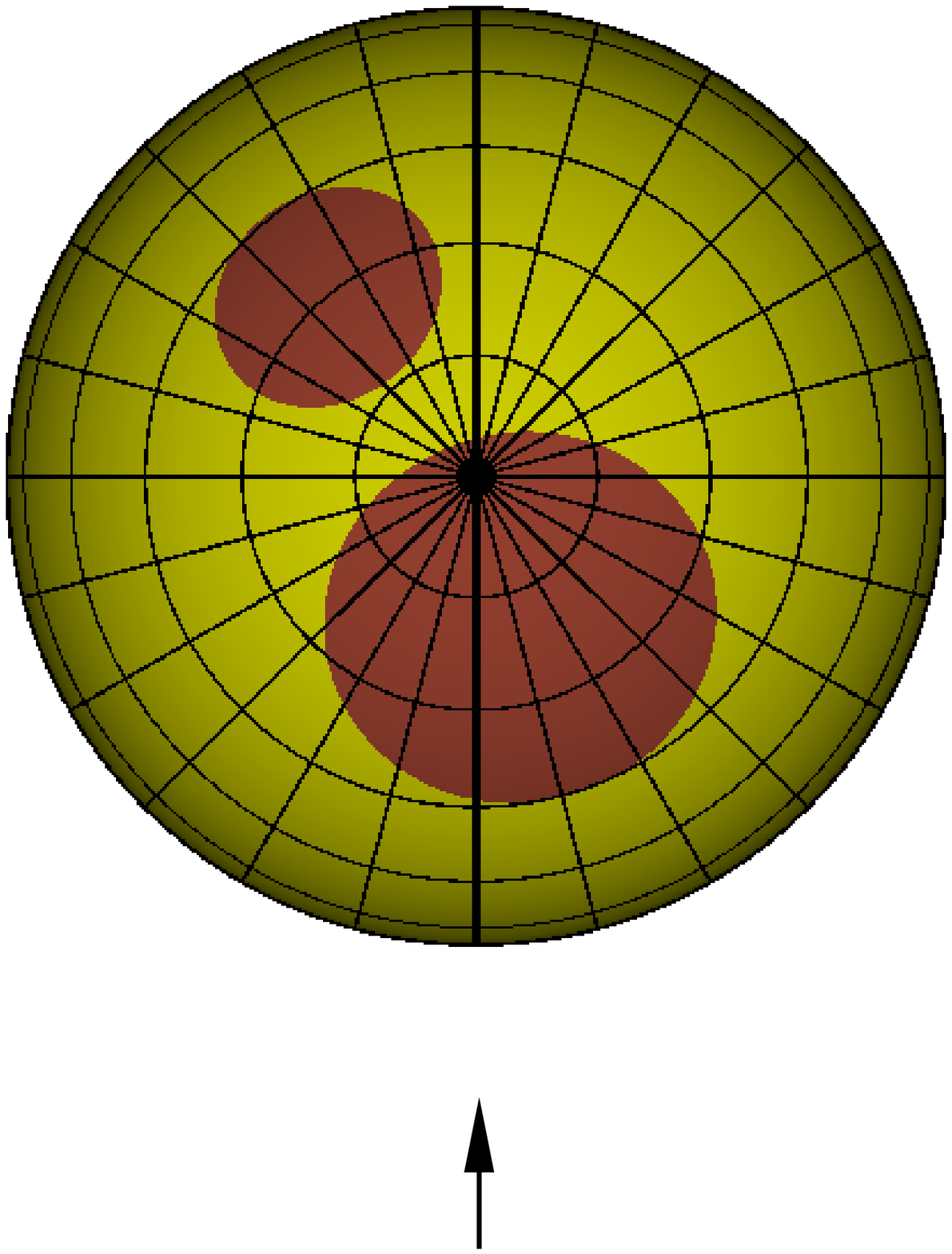}

 \caption{\textit{Left:} Phased $V$ light curve between JDs 2455029 and 2455036. The consecutive nights are plotted with different shades (getting lighter with time). The overall shape of the light curve remains the same, but nightly changes can be seen, especially at the location of the spot around 0.6 phase.
\textit{Right:} Spot model of V374 Peg from polar view. Arrow at the bottom indicates the line of sight at phase zero.
}
 \label{fig:nightly}
\end{figure}

When looking at the light curve (Fig. \ref{fig:all_v}) the first impression is that  -- except the flares -- there is no change in the mean brightness level, i.e. the total spottedness remains the same.
 A closer look reveals that during our observations two stable active regions could be found on the stellar surface: one at $\sim 0.0$ and another one at $\sim 0.6$ phase (see Fig. \ref{fig:phased}). 
For easier visualization we have made a spotmodel (see Fig. \ref{fig:nightly}) using the SML software \citep{sml} supposing a surface temperature of 3000K, a spot temperature of 2800K \citep{morin}, and circular spot shape. 
The overall shape of the phased light curve remains the same during our $\sim 430$ day-long time-series, which means, that these active regions do not change their size and position in appreciable extent.
Thus, from our  photometry we can confirm the finding of \cite{donati} and \cite{morin}: the light curve of V374 Peg is stable on the timescale of a year.

However, the light curve is not constant at all: small changes can be seen on a nightly timescale (see Fig. \ref{fig:nightly}). 
These variations are mainly seen around the active region at 0.6 phase. It is possible, that these variations are caused by newly-born and decaying spots in the same active region, similarly to the emerging flux ropes in active nests on the surface of the Sun.
We can conclude, that the location and size of the two active regions are stable, but the exact spot configuration is changing very rapidly.

Looking at Fig. \ref{fig:phased}, it can be clearly seen that V374 Peg has  active and less active states. During the first two observing runs frequent flaring could be observed, while in the next two runs the light curve is free of flares, with the spot configuration and the rapid nightly variations remaining the same as before (note, that there is a gap of $\sim 260$ days between the first two and the subsequent observing runs). During the following two runs the star is more active again, and in the last run the  phase coverage of the light curve is very poor, so no exact conclusions can be drawn from that one. It seems that the flaring and non-flaring phases on V374 Peg vary on monthly timescale.

The fact, that the nightly variation of the light curve occurs around the same phase ($\sim 0.6$) as the bulk of the flares, suggests that there is a connection between them. The frequent flares can be a result of reconnection between the emerging flux ropes of the active nest around phase 0.6. 

Our observations have raised some interesting questions. Is there any periodicity in the changing of the flaring/non-flaring state of V374 Peg? If yes, what is the underlying cause of this behaviour? Until now, all the studies have noted, that the magnetic configuration of the object seems to be stable. On what timescale are the large-scale changes happening? Hopefully these questions can be answered by further photometric monitoring and by more detailed modelling of the object.

\begin{acknowledgement}
 The authors are supported by the Hungarian Science Research Programme (OTKA) grants K-68626 and K-81421.
\end{acknowledgement}

%

\end{document}